**An Open-Source Web App for Creating and Scoring Qualtrics-based Implicit Association Test**

Yong Cui[1], Jason D. Robinson[1], Seokhun Kim[1], George Kypriotakis[1], Charles E. Green[1,2],

Sanjay S. Shete[3], Paul M. Cinciripini[1]

[1]Department of Behavioral Science, The University of Texas MD Anderson Cancer Center

[2]Department of Pediatrics, The University of Texas Health Science Center at Houston

[3]Department of Biostatistics, The University of Texas MD Anderson Cancer Center

Author Note: Correspondence regarding the article should be directed to Yong Cui, Department of Behavioral Science – Unit 1330, The University of Texas MD Anderson Cancer Center, P.O. Box 301439, Houston, Texas, 77030, the United States of America. Phone Number: +1-713-792-0919. Fax: +1-713-563-9760. Email: ycui1@mdanderson.org

**Abstract**

The Implicit Association Test (IAT) is a common behavioral paradigm to assess implicit attitudes in various research contexts. In recent years, researchers have sought to collect IAT data remotely using online applications. Compared to laboratory-based assessments, online IAT experiments have several advantages, including widespread administration outside of artificial (i.e., laboratory) environments. Use of survey-software platforms (e.g., Qualtrics) represents an innovative and cost-effective approach that allows researchers to prepare online IAT experiments without any programming expertise. However, there are some drawbacks with the existing survey-software as well as other online IAT preparation tools, such as limited mobile device compatibility and lack of helper functionalities for easy adaptation. To address these issues, we developed an open-source web app (GitHub page: https://github.com/ycui1-mda/qualtrics_iat) for creating mobile-compatible Qualtrics-based IAT experiments and scoring the collected responses. The present study demonstrates the key functionalities of this web app and describes feasibility data that were collected and scored using the app to show the tool's validity. We show that the web app provides a complete and easy-to-adapt toolset for researchers to construct Qualtrics-based IAT experiments and process the derived IAT data.

**Introduction**

Implicit attitudes are spontaneous and automatic evaluations that influence a variety of behavioral processes (Perugini, 2005), including racial stereotyping (McConnell and Leibold, 2001; Oswald et al., 2013), anxiousness and other personality characteristics (Schnabel et al., 2006; Grumm and Collani, 2007), and drug abuse (De Houwer et al., 2006; Waters et al., 2007). The most widely used behavioral paradigm to measure implicit attitudes is the Implicit Association Test (IAT; Greenwald et al., 1998), which measures reaction time to stimuli to evaluate the strength of associations between the target constructs under examination (e.g., insects vs. flowers) and the bipolar attribute concepts, such as valence (pleasant vs. unpleasant) and gender (male vs. female).

Conventionally, these IAT assessments are prepared and administered in a laboratory setting using desktop or laptop computer software (e.g., Inquisit, E-Prime). This computerized approach has the following limitations. First, it requires programming knowledge and considerable time commitment to create the task. Second, running these IAT experiments requires the installation and configuration of special software on the computers. Third, researchers can only run the IAT experiments locally, which precludes widespread deployment.

These limitations have led researchers to exploring options for running IAT experiments online, including Project Implicit (www.implicit.harvard.edu), Millisecond (www.millisecond.com), and Gorilla (Anwyl-Irvine et al., 2020). However, these online IAT deployment options are not entirely researcher-friendly for several reasons. For instance, most of them require considerable cost and excessive development time, particularly for custom IAT projects. Some tools also require test subjects install special software on their computers or subjects need to switch between online surveys and the websites that host the IAT experiments — these factors can lead to subject attrition (Wang-Jones et al., 2017).

An open-source R-based program recently developed by Carpenter and colleagues (2019) avoids many of these shortcomings by providing a web interface that allows researchers to generate IAT tests that run on the Qualtrics platform. One notable feature of Qualtrics is question customization with user-defined JavaScript code, which defines the appearance and data handling of the question. Taking advantage of Qualtrics' question customization feature, the web tool developed by Carpenter and colleagues can generate the needed JavaScript code for the IAT custom questions for stimulus presentation and response recording (Carpenter et al., 2019). This tool is innovative and useful as researchers can quickly prepare IAT tests without writing any code. More importantly, because the IAT assessments are deployed using the Qualtrics platform, any major web browser (e.g., Chrome, Firefox, Edge) is compatible, which not only eliminates the need of installing extra software on subjects' computers, but also makes the IAT tasks both locally and remotely accessible. As one of the most reputable online survey service providers, Qualtrics is the go-to survey platform for many researchers through their institutional subscriptions, and thus using Qualtrics entails little or no cost to researchers.

However, conducting Qualtrics-based IAT research with Carpenter and colleagues' tool has the following limitations. First, the generated IAT tests cannot be completed on mobile devices (e.g., smartphones, tablets). In terms of smartphones alone, the penetration rate of smartphones has been steadily growing in the last decade, and it is estimated that over 85% of adults in the United States own smartphones (Pew Research Center, 2021). Relatedly, web traffic through smartphone devices has exceeded 50% of the total traffic (Clement, 2021). Thus, the lack of mobile support may prevent many people from participating in IAT-related studies if they prefer mobile access.

Second, Carpenter and colleagues' tool does not have helper functionalities for preparing the image stimuli used in the IAT. Besides texts, images are common stimuli used in IAT experiments. To use images in Qualtrics-based IAT, researchers need to upload the images to the Qualtrics server (or other online file hosts) and retrieve the links to these images. These links are required to display the corresponding images in the IAT experiments. As of now, these processes can be done only manually by researchers, which are time consuming and error prone.

Third, adapting Carpenter and colleagues' tool to existing surveys is nontrivial. By design, the generated IAT uses 7 questions to represent each experimental block of the IAT, and thus the IAT survey consists of 28 questions to account for four counterbalanced presentation orders (crossing the two target categories and two attribute categories). Thus, integrating these many questions to an existing Qualtrics survey is nontrivial. In addition, when researchers need to make any modifications to the stimuli used in the IAT, they must make changes to each of these 28 questions, or they must re-generate the IAT survey template to create the survey again.

To address these issues, we developed an open-source web app for conducting Qualtrics-IAT research with the following enhancements. First, the generated IAT experiments can be run both on computers and mobile devices. Second, our web app provides automatic tools for researchers to upload images and download the links. Third, each IAT task is implemented as a single Qualtrics question in the survey such that it is easy for researchers to adapt our tool and make any necessary modifications. In this report, we describe the core functionalities of our web app and provide data collected from the flower-insect IAT created using our web app. The app's source code can be found at the project page: https://github.com/ycui1-mda/qualtrics_iat, where the web app's link is provided.

**The IAT Procedure**

In an IAT, participants classify stimuli (e.g., words and images) according to two complementary category labels. For illustration purposes in this report, we will use the flower-insect IAT to describe the details of the IAT procedure. In the flower-insect IAT, researchers use two sets of bipolar concepts: the target concept (flowers vs. insects) and the attribute concept (pleasant vs. unpleasant). The classic flower-insect IAT experiment consists of seven blocks with their respective classification labels and the number of trials shown in **Table 1**.

| Block | Concept Categories | Trials | Classification Labels (Left vs. Right) |
|-------|--------------------|--------|----------------------------------------|
| **Table 1.** Blocks of a representative flower-insect IAT experiment. | | | |
| 1 | Target | 20 | Flower vs. Insect |
| 2 | Attribute | 20 | Pleasant vs. Unpleasant |
| 3 | Target & Attribute | 20 | Flower or Pleasant vs. Insect or Unpleasant |
| 4 | Target & Attribute | 40 | Flower or Pleasant vs. Insect or Unpleasant |
| 5 | Attribute | 20 | Unpleasant vs. Pleasant |
| 6 | Target & Attribute | 20 | Flower or Unpleasant vs. Insect or Pleasant |
| 7 | Target & Attribute | 40 | Flower or Unpleasant vs. Insect or Pleasant |

**Note.** In some IAT versions, Block 5 reverses the order of the target labels instead of the attribute labels, as used in the example. Our web app allows the users to switch attributes or targets in Block 5.

The conditions shown in **Table 1** are for a representative flower-insect IAT experiment. However, to account for any order effect of these category labels (Greenwald et al., 2003) , there are four counterbalanced conditions (**Table 2**).

**Table 2.** The four permutations of block labels (left vs. right are shown).

| Permutation | Block 1 | Block 2 | Blocks 3 & 4 | Block 5 | Blocks 6 & 7 |
|-------------|---------|---------|--------------|---------|--------------|
| 1 | F vs. I | P vs. U | F or P vs. I or U | U vs. P | F or U vs. I or P |
| 2 | I vs. F | P vs. U | I or P vs. F or U | U vs. P | I or U vs. F or P |
| 3 | F vs. I | U vs. P | F or U vs. I or P | P vs. U | F or P vs. I or U |
| 4 | I vs. F | U vs. P | I or U vs. F or P | P vs. U | I or P vs. F or U |

**Note.** P: Pleasant, U: Unpleasant, F: Flower, and I: Insect. Block 5 reverses the attribute labels used in Block 2.

The premise for the IAT is that people react at different speeds in their classification of the presented stimuli as a function of the label pairings (Greenwald et al., 2002). When the pairings are in the expected association, people have faster reaction, while the pairings are unexpected, people have slower reaction. In the flower-insect IAT, the expected association is flower and pleasant, and such pairing is termed as compatible or congruent, while the opposite association (flower and unpleasant) is termed as incompatible or incongruent. The shorter reaction time of the congruent condition compared to that of the incongruent condition suggests a stronger association between the involved concepts (e.g., flower and pleasant).

To quantify the effect of the congruency on the reaction time, researchers compute the D scores (Greenwald et al., 2003), which are generally used as a measure of the IAT effect. The details of the algorithms used in computing the D scores will be discussed in the Scoring of the IAT Data section.

<p align="center">**Implementation of the IAT in Qualtrics**</p>

**Design Considerations**

Our goal with this tool was to minimize researchers' work of designing and implementing IAT assessments into research studies. Thus, our design accounts for the following factors. First, the web app provides an intuitive interface for researchers to configure all the parameters needed for creating IAT experiments. Second, the configuration creates a Qualtrics survey template file, from which researchers can directly create their IAT survey without any additional configurations. Third, we use a single question approach by customizing just one question as the vehicle for all IAT-related stimulus presentation and response recording. It will be easy to integrate the one-question-based IAT paradigm into other surveys.

**Configurable Parameters**

The most essential parameters are the target and attribute categories and their respective stimuli. Our web app supports the classic IAT experiment, which uses two target categories and two attribute categories. These four category labels must be specified. For each category, researchers provide the list of the stimuli. When the stimuli are words, researchers list the applicable words. When the stimuli are images, researchers list the links to these images. Besides these required parameters, there are several other optional parameters, which provide advanced configurations of the IAT experiment. Here, we highlight the parameters that are more likely to be configured than others.

**Study Name.** Researchers can specify the study name for the IAT experiment. This parameter will be used to name the embedded data fields in the Qualtrics survey, and these embedded data fields will record trial responses of the IAT experiment. This setting is necessary if researchers need to have multiple IAT studies in a single Qualtrics survey (more details are provided in the Multiple IAT Studies section).

**Reserve Attributes or Targets in Block 5.** The IAT experiment consists of compatible and incompatible conditions. Blocks 3 and 4 present one of these two conditions and Blocks 6 and 7 present the opposite condition. To create the opposite condition that is to be used in Blocks 6 and 7, Block 5 reverses the order of the attributes used in Block 2 or the targets used in Block 1. This setting will set whether the attributes or the targets to be switched in Block 5.

**Correction Requirement.** This parameter controls whether subjects are required to make a correction after making a wrong response. When not required, researchers can specify the delay for the task to automatically proceed after recording the wrong response.

**Colors for the Labels and Stimuli.** Researchers can optionally specify the colors for the classification labels. Target and attribute labels can be set to have different colors. Researchers can also specify the colors for the words presented for each category.

**Single Question Approach**

To simplify the adaptation, we take a single question approach — the IAT experiment needs just a single question, which holds the custom JavaScript code to present all the instructions and trials of all seven experimental blocks. The type of the question is Text/Graphic. There is no need to set any text or image to the question, which simply serves as a placeholder for controlling all HTML elements required for the IAT survey. The most essential functionalities included in the code are reviewed here.

**Stimulus Elements.** Before launching the IAT blocks, a series of HTML elements will be created and these elements are responsible for displaying the overall instruction, block-specific instructions, classification labels, the fixation cross, and the text or image stimuli. Based on the task's progress, the code will control the elements by showing relevant elements and hiding irrelevant elements.

**Stimuli Sources and Categories.** Both texts and images can be used as the stimuli. When the stimuli are texts, the stimuli should be the words themselves. However, when the stimuli are images, the stimuli should be the links to the images. These data will be added to the code automatically after the users specify them in the web app. Because the time for loading images can vary due to different factors (e.g., device specifications and internet speed), the code will download and cache the images before each block, which can help minimize the variation of image loading.

In the IAT experiment, there are four categories of stimuli: two for the targets and two for the attributes. For simplicity, in the code, we use four single characters to denote these categories: p = the positive target, n = the negative target, + = the positive attribute, and - = the negative attribute.

**Stimulus Randomization.** At the beginning of each block, the code will generate the needed stimuli for the block using the randomization without replacement procedure. In other words, the code will randomly draw stimuli from the proper categories until every stimulus is used in this round before the next round of drawing. For the combined blocks (i.e., Blocks 3, 4, 6, and 7), the target stimuli and the attribute stimuli alternate between consecutive trials.

**Task Randomization.** There are four possibilities for the classification labels used in the first combination block (Block 3): positive target + positive attribute, positive target + negative attribute, negative target + positive attribute, and negative target + negative attribute. Once the labels of this block are determined, the labels of the remaining blocks will be determined accordingly. The code can randomize the order between different surveys by randomly picking one of these four possibilities. Alternatively, the users can disable this randomization feature and implement counterbalancing by setting the proper survey flow, which will be discussed in the Survey Flow section.

**Handling of Incorrect Responses.** By default, when an incorrect response is entered, an error message (i.e., a red X) will be displayed. Researchers can require corrections be made. Alternatively, the task will move to the next trial after the specified delay (the default delay is 300 milliseconds when corrections are not required). In either case, the trial's response will be timestamped at the first attempt (i.e., the incorrect response) by default. However, it is possible to set the reaction time to be recorded until a correct response is entered when correction is required.

**Data Recording.** All the responses are saved as embedded data in the survey. Because Qualtrics has a size limit for the embedded data, to avoid any unexpected data loss, the JavaScript code will save all the trial data by block. The survey will record the stimuli (the word or the link to the image) for each block with stimuli separated by commas, and thus a block's stimuli can be saved as "Orchid,Tulip,Lilac..." with the embedded data field named as block1Trials.

Reaction time is recorded as the time difference between the launch of the stimulus and a response. Trial responses are saved in the format: trial number + correctness (Y or N) + reaction time in milliseconds, with trial responses separated by underscores. Thus, a block's trial responses are saved as "1Y673_2Y837_3N897..." with the embedded data field named as block1Responses.

**Mobile Compatibility**

Mobile compatibility of the IAT survey is achieved by implementing two features. First, the layout of the HTML elements is tailored, such as their size and position, based on the device type (computers vs. mobile devices). Second, participants' responses are recorded differently. For computers, we record the keystrokes on a keyboard, while for mobile devices, we record the tap events. Instead of having researchers or subjects specify the device information, the device type is automatically detected by the IAT survey.

**Survey Flow**

A typical IAT survey includes two required elements plus one optional element (**Figure 1**). The first required element is to set the embedded data, including a total of 15 fields: seven (e.g., block1Responses, block2Responses, ...) for saving trial responses, seven (e.g., block1Trials, block2Trials, ...) for saving stimuli, and one for saving the block conditions (blockConditions). Setting these embedded data fields are required to save any IAT data. The other required element is the IAT block, which consists of the single question that controls the entire IAT. The optional element is the randomizer, which determines the condition of the first combination block by setting a proper value to the firstCombination embedded field. Once this condition is decided, the conditions for all the other blocks will be determined accordingly.

The randomizer element is optional because the custom JavaScript code can provide randomization when the conditions are not set through the randomizer. However, the advantage of the Qualtrics' randomizer is that Qualtrics can track the assignments for each condition such that the users can achieve good counterbalancing between different IAT sessions. In other words, using the randomizer can guarantee that the conditions of the blocks are balanced for a study while code-implemented randomization is totally random, which has the chance of resulting in unbalanced assignment, particularly when the sample size is small.

**Figure 1.** The survey flow of the IAT Qualtrics survey

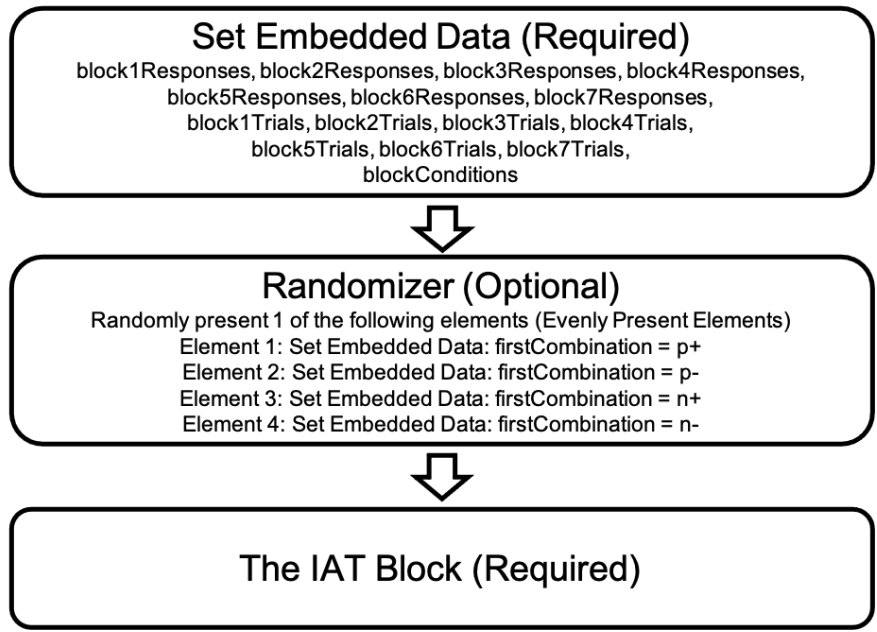

## Multiple IAT Studies

Because of our single question design, it is easy to have multiple IAT experiments in a single survey. To implement this, the following changes are necessary. Suppose that the user's Qualtrics survey includes the flower-insect and gender-career IAT experiments. First, each IAT experiment should have a unique study name (i.e., flower, gender), which the user can specify in the web app. Other than the specification of the IAT study name, all the remaining changes only involve updating the survey flow. Second, create the separate set of embedded data with each of the study name as the prefix. For instance, instead of block1Responses, the user should have flower_block1Responses and gender_block1Responses. Third, if the user elects to use the randomizer feature, the user should create separate randomizers for each task. In addition, the firstCombination fields should be named using the study name (i.e., flower_firstCombination in the flower IAT's randomizer and gender_firstCombination in the gender IAT's randomizer). Fourth, add each of the generated IAT experiment questions to the survey and add these two questions to the survey flow.

**IAT Scoring**

The goal of scoring the IAT data is to calculate the D score (Greenwald et al., 2003). Although scoring the IAT data can be done in any statistical program, it is a routine job that a shared scoring tool will benefit the researchers who use our web app. Therefore, the web app has an IAT scorer module, which implements two common IAT scoring algorithms: the conventional and the improved ones, developed by Greenwald and colleagues (Greenwald et al., 2003; see Table 3).

| **Table 3.** Algorithms for computing IAT scores | |
|---|---|
| **Conventional Algorithm** | **Improved Algorithm** |
| 1. Use data from B4 and B7<br>2. Exclude subjects with excessively slow responding (>3,000 ms) and/or high error rates (>10%)<br>3. Drop the first two trials of each block<br>4. Recode latencies outside the 300 – 3,000 ms using the boundary values<br>5. Log-transform the latencies<br>6. Average the resulting values for each block<br>7. Compute the difference between the means of B4 and B7 | 1. Use data from B3, B4, B6, and B7<br>2. Eliminate trials with latencies greater than 10,000 ms; Exclude subjects who have more than 10% of the trials with latency less than 300 ms<br>3. Use all trials (alternative: delete trials with latencies below 400 ms)<br>4. Compute the means of correct latencies for each block; Compute the SD of correct latencies for each block<br>5. Compute the pooled SD for all trials (alternative: correct only) in B3 and B6, and the pooled SD for all trials (alternative: correct only) in B4 and B7<br>6. Replace each error latency with block mean plus 600 ms (alternative 1: replace each error latency with block mean plus 2 x block SD of correct responses; alternative 2: use latencies to correct responses when correction to error responses is required, this alternative is only valid when the reaction time is recorded such)<br>7. Average the resulting values of each block<br>8. Compute two differences: B6 – B3 and B7 – B4<br>9. Divide each of these two differences by its associated pooled SD<br>10. Average the two quotients |

**Note.** Block is abbreviated as B (e.g., B3 means Block 3). ms: milliseconds; SD: standard deviations. For the applicable steps, alternatives are shown, and the web app provides corresponding options for these configurations.

To use the scoring tool, researchers will first upload the IAT data in the CSV format. The data file for the IAT survey should be downloaded from the Qualtrics server. Researchers can download the file manually or using the survey response downloading tool (discussed in the next section) of the app.

After the upload of the file, the web app will start to verify whether the IAT data has the correct format that is generated from the IAT survey and complete the initial cleanup of the IAT data, which will result in a dataset that consists of scored trial-level data, including the trial's block number, trial counter within the block, correctness, and reaction time. Researchers can download this data file as their own record, and they can further inspect these data for any potential issues or use a different scoring tool.

With the trial-level data, researchers can then specify which algorithm (conventional vs. improved) that they want to apply for scoring the IAT data. Once the algorithm is selected, the web app provides the list of related parameters with the default values set. Researchers can modify these parameters to other applicable values.

After the algorithm is selected with the parameters set, the web app can start the calculation. The output of the calculation consists of three parts. The first part is the summary of the algorithm with its parameters shown. The second part is the summary dataset, which provides descriptive information of the dataset, such as the number of IAT sessions scored, the number of trials exceeding the upper limit of the reaction time, the number of recoded error trials, the average D score, and reliability scores. The reliability score is computed by correlating two sets of D scores generated from the split-half odd and even trials and applying the Spearman-Brown correction (Li et al., 1996), and these procedures have been used in previous IAT studies (De Houwer and Bruycker, 2007; Carpenter et al., 2019). When the algorithm is the improved one, our app will also calculate a second reliability score by using the D scores generated from the Blocks 3 and 6 and Blocks 4 and 7, following the procedure in our early studies (Waters et al., 2007; Cui et al., 2019).

The third part will be the scored dataset with each data row representing an IAT session, and the key scored fields will include the D score, the reaction time for the congruent condition, and the reaction time for the incongruent condition.

## Qualtrics Tools

The web app has implemented five functionalities related to the Qualtrics platform. These functionalities are intended to replace some manual processes that researchers otherwise must carry out to implement the IAT survey. To use these functionalities, the users need to provide their API token, data center, and subdomain. These data are required to use Qualtrics APIs and the instruction for obtaining these values is provided in our web app.

### Upload Images

The users can upload images from their local computers to the Qualtrics server. They will specify the graphics library where these images will be uploaded to. A folder with the specified name will store these uploaded images. Importantly, the successful upload of images will result in the retrieval of the links to these images. These links can be used as stimuli in the IAT survey.

### Create Surveys

The web app can generate surveys programmatically for the users. They can upload a Qualtrics template file, such as the IAT template that the user creates from our tool. It should be noted that a Qualtrics template uses the JSON format. Thus, as an alternative, the users can specify JSON text directly to create the new survey.

### Export Survey Responses

The users can specify the ID number of the survey. The web app can export the responses for the survey in the formats of CSV or SPSS.

**Delete Images**

We expect that users may want to delete images from the Qualtrics server when these images are no longer needed. The users can specify the list of links or ID numbers for the images in the specific library.

**Delete Survey**

The users can delete a survey by specifying its ID number. To avoid deleting a survey accidentally, the users must enter "Delete Survey" to confirm the deletion of the survey.

## Empirical Study

**Study Overview**

We conducted an empirical study by recruiting participants from Amazon Mechanical Turk (referred to as MTurk) to complete a Qualtrics survey that included two IAT tests, including the flower-insect IAT and smoking IAT; the results of the latter will be reported separately. Participants would first be screened for eligibility. Only eligible participants were enrolled and proceeded to the IAT portion. After completing two IAT tests, they answered questions about basic demographics and be provided with a unique survey code. They submitted the code on the MTurk website to show their completion of the survey. After their responses were reviewed and approved, they received four dollars as their compensation.

The study was approved by the Institutional Review Board at The University of Texas MD Anderson Cancer Center. All participants provided consent to participate in the study at the beginning of the survey.

**Methods**

**Participants.** To be eligible for the study, participants met the following inclusion criteria: 1) 18 years old or above, 2) fluent in English, and 3) an MTurk Masters worker. Because we were interested in evaluating the device compatibility of the generated IAT, we recruited two batches of participants, with one requiring mobile devices and the other requiring computers to complete the IAT. Participants' device type was determined by the built-in tool available in the Qualtrics platform.

**The Flower-Insect IAT.** The bipolar attributes were pleasant and unpleasant, and the bipolar targets were flower and insect. The pleasant category included Joy, Happy, Laughter, Love, Friend, Pleasure, Peace, and Wonderful. The unpleasant category included Evil, Agony, Awful, Nasty, Terrible, Horrible, Failure, and War. The flower category included Orchid, Tulip, Rose, Daffodil, Daisy, Lilac, and Lily. The insect category included Wasp, Flea, Roach, Centipede, Moth, Bedbug, and Gnat. There were seven blocks: the first (20 trials) was to classify the targets (flower vs. insect), the second (20 trials) the attributes (pleasant vs. unpleasant), the third (20 trials) and fourth (40 trials) the targets and attributes using the combined conditions (flower or pleasant vs. insect or unpleasant), the fifth (20 trials) the attributes using the reverse order (unpleasant vs. pleasant), and the sixth (20 trials) and seventh (40 trials) using the combined conditions (flower or unpleasant vs. insect or pleasant). We used the Qualtrics randomizer to counterbalance the four possible combinations of the conditions.

**Data Collection.** We published two batches of tasks on MTurk, with one batch requiring mobile devices to complete and the other computers. All survey data were saved automatically on the Qualtrics server after participants completed the survey.

**Data Scoring.** We used the scorer tool within the web app. For the current report, we used the improved algorithm with the following parameters (Greenwald et al., 2003). The high cutoff for the reaction time to be considered a slow responding was 10,000 ms. The trials exceeding the 10,000 ms cutoff were eliminated from further steps. A trial with less than 300 ms response time was considered a fast response. Subjects with over 10% of their trials as fast responses were excluded. All trials in Blocks 3 and 6 were used to calculate the pooled SD, and the same procedure applied to Blocks 4 and 7. For the error trials, their response time latencies were replaced with the mean reaction time of the correct trials of the block plus 600 ms punishment time.

**Statistics.** Data processing and analyses were all performed with Python and its packages, including pandas (McKinney, 2011) and scipy (Virtanen et al., 2020). A one-sample t-test was used to examine the effect of the congruency of the label pairings on the D score. Because the D score was already a difference score between the incongruent and congruent condition, the null hypothesis for the t-test was that the D score was 0, meaning that there was no difference in the reaction time between the congruent and incongruent conditions. We calculated Cohen's d (Cohen, 1988) to estimate the effect size of the IAT D score. To examine the possible effect of the device type on the IAT effect, we conducted a series of two-sample independent t-tests. The significant level was 0.05.

**Results**

**Sample Characteristics.** Our survey received 134 responses in total, including 100 responses as complete MTurk submissions across two MTurk batches. Among the remaining 34 responses, only 1 response was complete, but the responder did not submit the survey code on the MTurk website. The other 33 were partial responses: 29 due to using a computer when a mobile device was required, 1 due to being ineligible, and 3 were not completed. Our analysis used the 100 responses through complete MTurk submissions, which represented 88 unique MTurk workers. Their demographics are shown in **Table 4**.

**Table 4.** Demographics of the Sample (N = 88)

| Variable | N (%) |
| --- | --- |
| **Gender** | |
| Male | 30 (34.1) |
| Female | 58 (65.9) |
| **Age in years** | |
| 18 – 30 | 14 (15.9) |
| 31 – 43 | 53 (60.2) |
| 44 and above | 21 (23.9) |

| **Education** | |
| --- | --- |
| High school | 33 (37.5) |
| Bachelor's degree | 47 (53.4) |
| Master's degree | 6 (6.8) |
| Doctoral degree | 1 (1.1) |
| Prefer not to say | 1 (1.1) |
| **Race/Ethnicity** | |
| Asian or Pacific Islander | 15 (17.0) |
| Black or African American | 1 (1.1) |
| Hispanic or Latino | 2 (2.3) |
| White | 68 (77.3) |
| Other | 2 (2.3) |

**Behavioral Performance.** The average error rate of the 100 responses was 11.35% (SD: 10.48%, range: 0.56 – 52.78%). The overall error rate is comparable to the error rates (8 – 10%) found in the IAT studies conducted among the MTurk workers (Carpenter et al., 2019). Only 8 trials had a response time over 10 s, which represented less than 0.05% of the total trials. 131 trials had a response time less than 300 ms, which represented 0.73% of the total trials. No subjects had more than 10% of the trials less than the 300 ms cutoff.

**The IAT Effect.** In the flower-insect IAT, we considered the congruent condition when the flower and pleasant labels were combined and the incongruent condition when the flower and unpleasant labels were combined. **Figure 2** shows the reaction time of the congruent and incongruent conditions. The reaction time of the incongruent condition was significantly longer than that of the congruent condition (t-value=11.87, p<0.0001), suggesting that people were more likely to associate flowers with the pleasant attribute than with the unpleasant.

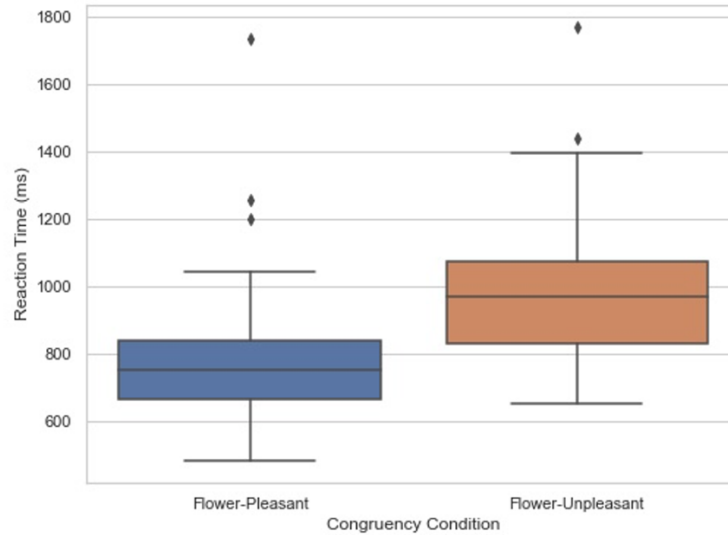

**Figure 2.** Longer reaction time in the flower-unpleasant condition than in the flower-pleasant condition in the flower-insect IAT.

We then calculated D score for each participant, and a positive D score suggested a stronger association between flowers and the pleasant attribute. We found that the average D score was 0.86 (SD=0.51). One-sample t-test revealed that the D score was significantly higher than 0 (t-value=16.7, p < 0.0001) with an estimated effect size of Cohen's d of 1.69.

However, the reliability of the IAT measure was modest. The reliability score using the split odd-even trials was 0.54, while the reliability score using the D scores generated from Blocks 3 + 6 and Blocks 4 + 7 was 0.61.

**The Impact of the Device Type.** We found that the D score of the IAT responses completed on mobile devices was not different from that of those completed on computers (t-value=0.46, p=0.64). In a similar fashion, the reaction time difference between the incongruent and congruent conditions for the mobile devices was not different from that for the computers (t-value=0.87, p=0.38). However, the reaction time of the responses completed on mobile devices was significantly longer than that on computers both for the congruent (t-value=3.93, p<0.001) and incongruent (t-value=4.11, p<0.001) conditions.

Because 12 MTurk workers participated in both batches, we used repeated measures to examine the impact of the device type as a within process. We found that they had faster reaction when they used computers than when they used mobile devices (congruent: t-value=4.08, p<0.005; incongruent: t-value=3.94, p<0.005). As expected, the D scores and reaction time differences between the incongruent and congruent conditions between these two device types were not different (D score: 0.91 vs. 0.99, t-value=0.40, p=0.70; reaction time difference: 153.62 vs. 175.71 ms, t-value=0.75, p=0.47).

**Discussion**

The IAT method is a well-validated experimental paradigm to examine implicit attitudes in various psychological research domains (Lane et al., 2007). Administering IAT experiments online has the advantages of collecting data remotely and having access to a wide audience, and both aspects will facilitate research progress. However, there are a variety of difficulties in running online experiments, such as considerable cost, development time, and technical challenges. In this article, we describe a free open-source web app that allows researchers to generate IAT experiments that can run on the Qualtrics platform. In addition, our web app provides functionalities for scoring the IAT responses generated from the Qualtrics survey and for interacting with the Qualtrics server (e.g., image uploading and response downloading). We also ran a feasibility IAT study using the described web app.

Compared to other online IAT experiment preparation tools, there are several advantages of our tool. First, it requires relatively little time and cost for researchers to develop a fully functioning IAT experiment running on the Qualtrics platform. Researchers can configure all the essential parameters for the IAT experiment to produce a Qualtrics template file. From this template, researchers can create a Qualtrics survey either using the automatic survey creation tool in our app or manually using the Qualtrics website. Importantly, the survey is ready to be distributed to their participants by the researchers because the generated template file has already configured the survey flow (e.g., setting the needed embedded data fields).

Second, the generated IAT Qualtrics surveys are compatible with mobile devices. The existing tool for developing Qualtrics-based IAT experiments lacks support of running IAT experiments on mobile devices, which prevents data collection from participants who prefer using mobile devices. Although other IAT experiment preparation tools, such as Gorilla (Anwyl-Irvine et al., 2020) and Project Implicit (www.implicit.harvard.edu), support mobile compatibility, there are different issues, including considerable development cost, incomplete support of mobile devices with different screen sizes, and the manual specification of mobile mode by users. By contrast, our IAT surveys can detect the device type and apply the proper HTML elements accordingly with the automatic justification of the device's screen size.

Related to the device type, it should be noted that we found in our feasibility study that people had longer reaction time among those who used mobile devices than those who used computers, reflecting that it took less time to press a key than touch the screen on a mobile device. This difference does not matter for the IAT paradigm, which is a within-subject measure by design. The differential behavioral reactions between subjects, no matter on a mobile device or a computer, would be cancelled out because the calculations involved only each subject's own performance in his/her conditions. However, other behavioral tasks whose performance is not measured in a within-subject manner should not overlook this inherent response time difference between mobile devices and computers.

Third, it is easy to prepare a Qualtrics survey that consists of multiple IAT experiments. By design, the implementation of our IAT experiment requires just a single question, such that when researchers need to have multiple IAT experiments included in a survey, they will need to deal with a limited number of questions. The only required change is the update of the survey flow, which takes little time if researchers follow the instruction as outlined in this report.

Fourth, our app provides the tool for scoring the generated Qualtrics survey responses. The generated survey responses have a special format that is specific to the Qualtrics platform. It will require considerable efforts to score these data. Because IAT data scoring is a routine task, our app provides the scoring functionality. Notably, we implement two algorithms, the conventional and the improved. The app has all related parameters configurable such that researchers can customize the scoring algorithms to meet their specific needs.

Fifth, our app supports the preparation of the image stimuli used in the IAT survey. It is tedious to retrieve the links to images from the Qualtrics server if this process is manual. Our app can automate this process by providing an image uploader, which will update the selected images from the local computer and retrieve the links for all the uploaded images.

Despite these advantages, our tool has the following limitations. First, it only supports the classic IAT experiments. For instance, the single target IAT (Karpinski and Steinman, 2006) is not directly supported by our app. However, researchers who have coding experience can use our code as a basis for an easy adaptation. Second, it is only integrable with the Qualtrics platform. Although Qualtrics is one of the most popular survey platforms, some researchers may not have access to this platform through their institutions. Thus, researchers may have to register a Qualtrics account before they can use our tool. Third, the reliability of the IAT score from the empirical study was modest. We speculated that it was mainly due to the sample used in this study (i.e., MTurk workers), because in one of Carpenter's IAT studies conducted among MTurk workers (Carpenter et al., 2019), they observed a similar low consistency. It is possible that the inadequate performance of MTurk workers may have compromised the consistency of the IAT measure. Nevertheless, future studies need to be conducted to compare the reliability of these tools.

To conclude, our web app is an open-source tool for researchers to conveniently conduct Qualtrics-based IAT experiments. Through our app, researchers can prepare the IAT stimuli (e.g., image links), create the IAT survey template, deploy the survey, and analyze the survey responses. We expect that our web app will facilitate IAT-related research as an open-source tool, and we look forward to any contribution to this open-source project from IAT researchers.

## Funding

Work on this project was supported by MD Anderson's Cancer Center Support Grant (P30CA016672) from the National Cancer Institute.

## Conflict of Interest

The authors declare no conflict of interest.

## Acknowledgement

The authors wish to thank Ms. Jennifer Ferguson and Ms. Carmen Pollock for their help in conducting the research project.

## Open Practices Statement

The data and materials for all experiments are available at the project's GitHub page.

## References

Anwyl-Irvine, A.L., Massonnié, J., Flitton, A., Kirkham, N., Evershed, J.K., 2020. Gorilla in our midst: An online behavioral experiment builder. Behavior Research Methods, 52 (1), 388–407.

Carpenter, T.P., Pogacar, R., Pullig, C., Kouril, M., Aguilar, S., LaBouff, J., Isenberg, N., Chakroff, A., 2019. Survey-software implicit association tests: A methodological and empirical analysis. Behav Res Methods, 51 (5), 2194–2208. doi:10.3758/s13428-019-01293-3.

Clement, J., 2021. Percentage of mobile device website traffic worldwide from 1st quarter 2015 to 1st quarter 2021. statista. https://www.statista.com/statistics/277125/share-of-website-traffic-coming-from-mobile-devices/. Accessed 13 July 2021.

Cohen, J., 1988. Statistical Power Analysis for the Behavioral Sciences, 2nd edn., Lawrence Earlbaum Associates, Hillsdale, NJ.

Cui, Y., Engelmann, J.M., Gilbert, D.G., Waters, A.J., Cinciripini, P.M., Robinson, J.D., 2019. The impact of nicotine dose and instructed dose on smokers' implicit attitudes to smoking cues: An ERP study. Psychol Addict Behav, 33 (8), 710.

De Houwer, J., Bruycker, E. de, 2007. The implicit association test outperforms the extrinsic affective Simon task as an implicit measure of inter-individual differences in attitudes. The British journal of social psychology, 46 (2), 401–421.

De Houwer, J., Custers, R., De Clercq, A., 2006. Do smokers have a negative implicit attitude toward smoking? Cogn Emot, 20 (8), 1274–1284.

Greenwald, A.G., Banaji, M.R., Rudman, L.A., Farnham, S.D., Nosek, B.A., Mellott, D.S., 2002. A unified theory of implicit attitudes, stereotypes, self-esteem, and self-concept. Psychol Rev, 109 (1), 3–25.

Greenwald, A.G., McGhee, D.E., Schwartz, J.L.K., 1998. Measuring individual differences in implicit cognition: The implicit association test. J Pers Soc Psychol, 74 (6), 1464–1480. doi:10.1037/0022-3514.74.6.1464.

Greenwald, A.G., Nosek, B.A., Banaji, M.R., 2003. Understanding and using the implicit association test: I. An improved scoring algorithm. J Pers Soc Psychol, 85 (2), 197–216. doi:10.1037/0022-3514.85.2.197.

Grumm, M., Collani, G. von, 2007. Measuring Big-Five personality dimensions with the implicit association test–Implicit personality traits or self-esteem? Pers Indiv Differ, 43 (8), 2205–2217.

Karpinski, A., Steinman, R.B., 2006. The single category implicit association test as a measure of implicit social cognition. J Pers Soc Psychol, 91 (1), 16.

Lane, K.A., Banaji, M.R., Nosek, B.A., Greenwald, A.G., 2007. Understanding and using the Implicit Association Test: IV: What we know (so far) about the method. (None).

Li, H., Rosenthal, R., Rubin, D.B., 1996. Reliability of measurement in psychology: From Spearman-Brown to maximal reliability. Psychol Methods, 1 (1), 98.

McConnell, A.R., Leibold, J.M., 2001. Relations among the Implicit Association Test, discriminatory behavior, and explicit measures of racial attitudes. Journal of experimental Social psychology, 37 (5), 435–442.

McKinney, W., 2011. pandas: A foundational Python library for data analysis and statistics. Python for high performance and scientific computing, 14 (9), 1–9.

Oswald, F.L., Mitchell, G., Blanton, H., Jaccard, J., Tetlock, P.E., 2013. Predicting ethnic and racial discrimination: A meta-analysis of IAT criterion studies. J Pers Soc Psychol, 105 (2), 171.

Perugini, M., 2005. Predictive models of implicit and explicit attitudes. The British journal of social psychology, 44 (Pt 1), 29–45. doi:10.1348/014466604X23491.

Pew Research Center, 2021. Mobile Fact Sheet. https://www.pewresearch.org/internet/fact-sheet/mobile/. Accessed 4 June 2021.

Schnabel, K., Banse, R., Asendorpf, J.B., 2006. Assessment of implicit personality self-concept using the implicit association test (IAT): Concurrent assessment of anxiousness and angriness. The British journal of social psychology, 45 (2), 373–396.

Virtanen, P., Gommers, R., Oliphant, T.E., Haberland, M., Reddy, T., Cournapeau, D., Burovski, E., Peterson, P., Weckesser, W., Bright, J., 2020. SciPy 1.0: Fundamental algorithms for scientific computing in Python. Nature methods, 17 (3), 261–272.

Wang-Jones, T., Alhassoon, O.M., Hattrup, K., Ferdman, B.M., Lowman, R.L., 2017. Development of gender identity implicit association tests to assess attitudes toward transmen and transwomen. Psychology of Sexual Orientation and Gender Diversity, 4 (2), 169.

Waters, A.J., Carter, B.L., Robinson, J.D., Wetter, D.W., Lam, C.Y., Cinciripini, P.M., 2007. Implicit attitudes to smoking are associated with craving and dependence. Drug Alcohol Depend, 91, 178–186.